# SPEAKER DETECTION IN THE WILD: LESSONS LEARNED FROM JSALT 2019

*Paola Garcia*[1], *Jesús Villalba*[1], *Hervé Bredin*[2], *Jun Du*[3], *Diego Castan*[4], *Alejandrina Cristia*[5],
*Latané Bullock*[9], *Ling Guo*[8], *Koji Okabe*[8], *Phani Sankar Nidadavolu*[1], *Saurabh Kataria*[1], *Sizhu Chen*[10],
*Leo Galmant*[2], *Marvin Lavechin*[5], *Lei Sun*[3], *Marie-Philippe Gill*[7], *Bar Ben-Yair*[1], *Sajjad Abdoli*[7], *Xin Wang*[3],
*Wassim Bouaziz*[5], *Hadrien Titeux*[5], *Emmanuel Dupoux*[6], *Kong Aik Lee*[8], *Najim Dehak*[1]  *

[1] Center for Language and Speech Processing, The Johns Hopkins University, Baltimore, USA,
[2] LIMSI, CNRS, Univ. Paris-Sud, Université Paris-Saclay, Orsay, France
[3] University of Science and Technology of China, Hefei, China
[4] Speech Technology and Research Laboratory, SRI International, California, USA
[5] Cognitive Machine Learning, ENS/INRIA/PSL, PSL Research University, Paris, France
[6] Cognitive Machine Learning, ENS/CNRS/EHESS/INRIA/PSL, PSL Research University, Paris, France
[7] École de Technologie Supérieure, Université du Québec, Montreal, Canada
[8] NEC Corporation, Japan, [9] Rice University, Houston, USA, [10] University of California, San Diego

leibny@gmail.com

## ABSTRACT

This paper presents the problems and solutions addressed at the JSALT workshop when using a single microphone for speaker detection in adverse scenarios. The main focus was to tackle a wide range of conditions that go from meetings to wild speech. We describe the research threads we explored and a set of modules that was successful for these scenarios. The ultimate goal was to explore speaker detection; but our first finding was that an effective diarization improves detection, and not having a diarization stage impoverishes the performance. All the different configurations of our research agree on this fact and follow a main backbone that includes diarization as a previous stage. With this backbone, we analyzed the following problems: voice activity detection, how to deal with noisy signals, domain mismatch, how to improve the clustering; and the overall impact of previous stages in the final speaker detection. In this paper, we show partial results for speaker diarizarion to have a better understanding of the problem and we present the final results for speaker detection.

## 1. INTRODUCTION

For the past years, speaker recognition research has mainly focused on telephone and close-talk microphone applications with high speech quality levels. However, far-field recordings have recently emerged as an area of interest. This is due to new applications like video annotation; home assistant devices which need to distinguish between family members; and wearables that document our everyday life. These applications provide a massive amount of data which requires automatic means of analysis. Furthermore, such devices are often used in very challenging environments with multiple speakers, and where the audio is affected by noise and reverberation. Most devices use a single microphone and, therefore, multichannel signal processing techniques (e.g., beamforming) cannot be applied to alleviate the impact of the real-life conditions. As reference, speaker detection error rates of reverberant speech are 2 times worse than close-talk speech in voices dataset [1]; internet videos with noise and multi-speaker error multiplied by 6 in recent NIST SRE18 w.r.t. clean videos [2]; and systematic evaluation of diarization (Dihard I and Dihard II [3, 4]) in real-life domain shows diarization error rates above 60% when all the stages are automatized [5].

Knowing that speaker diarization and detection require further research, the aim of our workshop was to ivestigate, develop, and benchmark speaker diarization and speaker recognition systems on far-field speech using single microphones in realistic scenarios that include background noises The key aspects that we found were fundamental to investigate are: voice activity detection, speech enhancement, domain adaptation and improving clustering. Each part will be explained briefly in the following paragraphs.

## 2. RESEARCH THREADS

### 2.1. Initial system: baseline

Our baseline follows a traditional diarization workflow connected to a speaker detection branch (see Figure 1, solid line rectangles). For diarization (who speaks when in a recording), the first stage is a voice activity detection (VAD), followed by an acoustic feature extractor and an embedding extractor. The embedding extractor uses a sliding window to produce a sequence of speaker embeddings. The clustering block intends to group the sequence of embeddings into single speaker clusters. A speaker label is created for each cluster. Thus, the diarization output is a sequence of speaker labels with its corresponding time marks. The speaker detection stage performs a verification task, *i.e.*, decide whether a known speaker is in the recording or not. We are interested in an scenario where multiple speakers can be present in the test recording; therefore, we apply diarization as a first step. The diarization output is used to compute a speaker embedding for each of the speakers identified in the diarization stage. Then each test speaker embedding is compared against the enrollment embedding. To compare two embeddings we usually employ probabilistic linear discriminant analysis (PLDA), that outputs scores. When we encounter a mismatch condition those scores should need calibration. The metric employed for diarization is diarization error rate (DER), which takes into account the false alarms, the misses (classifying speech as non-speech) and speaker confusion. The speaker detection metrics are the EER, minimum and actual DCF.

To explore in detail the challenging conditions in far-field, we analyzed the shaded rectangles blocks in Figure 1.

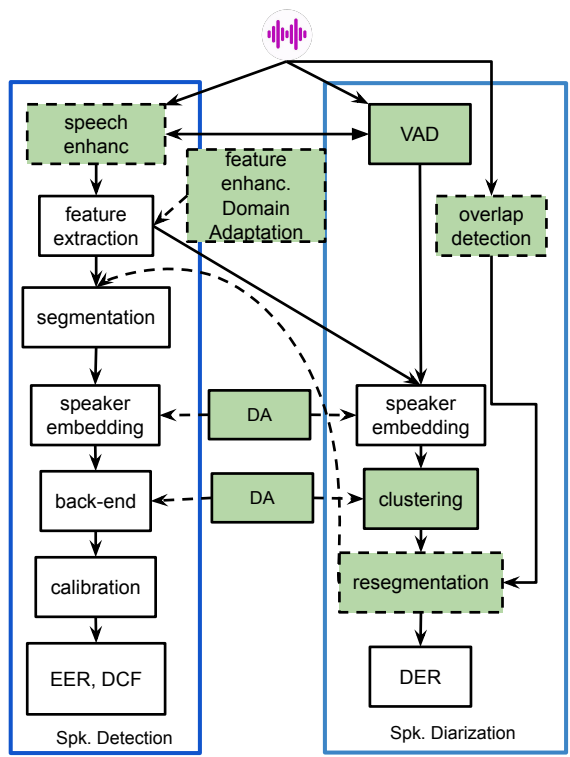

**Fig. 1**. Speaker detection pipeline. □ is the baseline system, ⬚ are the new parts that are included in the pipeline, shaded-□ are the modules that were investigated.

### 2.2. Voice Activity Detection and Speaker Separation

Improving speaker separation and voice activity detection addresses some of the key issues we had identified as being particularly problematic for our task. In this area, we explored three main routes.

The first one focuses on separating child and adult speech in realistic conditions [6]. First, the method measures the speech dissimilarities between children, babies and adults using i-vectors. Then, it uses a model based on progressive learning to extract child speech from simulated mixtures during training. The data from toddlers between 2 and 5 years is limited. Based on this constraint, a progressive learning framework generates intermediate target outputs and stack them together with the original limited-sized mixed input feature vectors to increase the training samples. To boost the speech separation we employed Multiple-target learning.

The second study addresses the *overlap detection* [7]. We hypothesized that detecting regions of overlapped speech (two or more speakers at the same time) is most effectively solved from the raw waveform. This approach relies on two stacked bi-LSTM layers, two feed-forward layers, and a final classification layer, fed into binary cross-entropy loss. The input can be MFCCs or Sincnet [8].

The third approach combines the VAD and speaker type classification in a multi-task learning set-up. We trained a system to diarize five classes: key child (wearing a recorder), other child, male adult, and female adult and speech, all of which could be active at the same time. Silence is marked by the absence of activation in the classes. The best architecture consisted of learning from the raw waveform with the SincNet model followed by a LSTM and a fully connected layer.

### 2.3. Dealing with noisy signals

We explored two main ideas to overcome the noisy signals in our different scenarios.

In the first approach, we built a SNR-progressive multi-target learning based *speech enhancement* model for adverse acoustic environments. The progressive multi-targets (PMT) network is divided into successively stacking blocks with one LSTM layer and one fully connected layer via multi-targets learning per block. The fully connected layer in every block (target layer) is designed to learn intermediate speech targets with a higher SNR than the targets of previous target layers. A series of progressive ratio masks (PRM) are concatenated with the progressively enhanced log-power spectra (PELPS) features together as the learning targets. At test time, we directly feed the enhanced audios processed by our enhancement model to the back-end systems, including speaker diarization and speaker detection. More details about this approach are described in [9].

The second approach we explored is *feature enhancement* with deep feature losses [10]. The main idea is to train a feature-domain enhancer which can serve as a pre-processing module to the x-vector system during inference. We develop on the ideas of perceptual loss [11] and speech denoising work [12]. This approach requires a pre-trained auxiliary network for loss estimation between enhanced and clean samples. For the auxiliary network, we chose an x-vector network based on ResNet-34 with LDE pooling and trained for speaker classification using an Angular Softmax loss objective [13]. For the enhancement network, we design networks based on [14] (Encoder-Decoder residual network) and [15] (Context Aggregation Network). We obtained significant improvement on real datasets using auxiliary x-vector network trained on clean speech. Using the augmented x-vector auxiliary network, we observed slight improvements on simulated noisy sets.

### 2.4. Domain Mismatch

We explored two ways to deal with domain mismatch between training and test data: domain adaptation in acoustic feature space and domain adversarial training for VAD training.

In our first case, we examined how to train an unsupervised speech enhancement system, which can be used as a front-end pre-processing module to improve the quality of the features before they are forward passed through the x-vector network. The details of the procedure can be found in [16]. Simply put, the unsupervised adaptation system is based on CycleGAN [17, 14]. We trained a CycleGAN network using log-Mel filterbanks as input to each of the generator network. During testing, we process the far-field test data through the *reverb to clean generator* of CycleGAN. These enhanced acoustic features are then used to extract x-vectors. Though CycleGAN network was trained for doing de-reverberation task, we also tested it on noisy datasets to investigate its generalization abiliy to unseen test conditions. We observed improvements on both reverberant and noisy test datasets.

We also proposed domain-adversarial training for robust end-to-end VAD [18]. On one branch an LSTM based end-to-end VAD labels segments as speech or non-speech based. On a second branch, a neural SincNet extracts filters that are the input to a domain-adversarial multitask training. Those filters via the adversarial process are forced to be domain independent. Each branch can perform its task independently, or they can be combined using gradient reversal that give robustness to the VAD.

### 2.5. Improving clustering

Knowing that *clustering* is a crucial stage in diarization, we examined a set of ways to address the problem.

A first solution was to turn the problem into a supervised learning task so that it considers the temporal information and optimization of the DER as the main goal. The model input is the set of embeddings from a recording with multiple speakers. This is an encoder-decoder model, where the encoder converts the sequence of embeddings into a context vector **c**. This vector contains information about the dialogue, like the number of clusters (speakers), cluster centroids, etc. The decoder compares embeddings from the input with a context vector and assigns a cluster. Both encoder and decoder used a bi-directional multi-layer RNN to deal with sequentiality.

A second approach was to refine the clustering by combining the *resegmentation with overlap detection* (see Section 2.2) [7]. The state-of-the-art method for the refinement is variational Bayes (VB) HMM (Hidden Markov Model) resegmentation, studied in [19]. VB-HMM resegmentation computes a per-frame speaker posterior matrix **Q**. Meanwhile, overlap detection provides the regions with two or more speakers talking. We hypothesized two speaker labels in those regions, *i.e.*, take the two speaker with higher posterior in matrix **Q** and continue the diarization process. The DER was reduced considerably for all the cases by including the overlap assignment.

### 2.6. Speaker detection

We explore how to overcome the main sources of confusion in speaker detection: overlap speech and the multiple speakers in the same audio stream. Firstly, we compare two different techniques to create homogeneous segments, computing the diarization as input to speaker detection and the sliding-window without knowledge of the speaker labels. Secondly, we used the overlap regions (explained in Section 2.2) to remove them from the speaker detection pipeline. The main conclusion of our experiments is that diarization is still necessary for these type of environments, where the speech of the speaker-of-interest is highly degraded by other sources of audio. While there is a slight improvement in matching scenarios, the speaker detector degrades dramatically when the overlap detector is not trained within the same domain showing a clear problem with the purification in mismatch data.

## 3. EXPERIMENTAL SETUP

### 3.1. Datasets

To study the main issues of the far-field scenario we built a dataset that is summarized by the following four corpora:

- **Meeting** ( *AMI* [20]): with a setting of 3 different meeting rooms with 4 individual headset Microphones, 8 Multiple Distant Microphones forming a microphone array; 180 speakers x 3.5 sessions per speaker (sps); suitable for diarization and detection. Since we are exploring single microphones, we focused only on the mix Headset scenario.
- **Indoor controlled** ( *SRI data* [21] [1] ): with a setting of 23 different microphones placed throughout 4 different rooms; controlled backgrounds, 30 speakers x 2 sessions and 40 h, live speech along with background noises (TV, radio); suitable for detection (only reliable labeling of target speaker was provided).
- **Indoor not controlled** ( *CHiME5* [22] ): with a setting of kitchen, dining, living room, 80 speakers, 50 h; 4 speakers in two-hour recordings; 32 microphones per session; suitable for diarization only (there are not enough impostor speakers in different sessions within the corpus)
- **Wild** ( *BabyTrain*[2] ): with an uncontrolled setting, 450 recurrent speakers, up to 40 sps (longitudinal), 225 h; suitable for diarization and detection.

For *speaker detection* the enrollments were generated by accumulating non-overlapping speech ( 5, 15 and 30s duration) of every target speaker along one or multiple utterances. For the test, we cut the audio into 60 second chunks. We do a Cartesian product between the enrollments and the test segments to generate all possible trials. Then based on conditions, some trials are filtered out. For example, same session and same microphones are not allowed to produce a target-trial pair.

### 3.2. System configuration

Our experimental setup is depicted in Figure 1. The dotted blocks are the new approaches added to the pipeline. The underlined methods along the text show the final combination of modules that obtained effective results and that are part of the final code contribution in [24]. The definite pipeline follows:

1. **Speech enhancement(Diar/SpkD)**: we used a 1000-hour training set. The noisy mixtures are made at three SNR levels (-5dB, 0dB and 5dB), and the progressive increasing SNR between two adjacent targets is set to 10 dB. The audios are sampled at 16 kHz rate and the frame length is 256 samples (PELPS and the PRM are 257 dimension). During the testing stage, we directly feed the enhanced audios processed by our enhancement model to back-end systems.
2. **Acoustic features(Diar/SpkD)**: 23 dimension MFCC for x-vector systems based on Kaldi TDNN x-vectors; and 23 log-Mel filter banks for ResNet based x-vectors. Features were short-time centered before silence removal with a 3 seconds sliding window.
3. **Feature enhancement(Diar/SpkD)**: We use a SNR estimation to select the 50% highest SNR signals of VoxCeleb [25]

---

[1] This data was recorded by SRI international and was submitted to LDC for publication

[2] This data uses daylong recordings from Homebank [23], expected to be public

|  | AMI | | | | BabyTrain | | | | CHiME5 | | | |
|---|---|---|---|---|---|---|---|---|---|---|---|---|
|  | DER | FA | Miss | Conf | DER | FA | Miss | Conf | DER | FA | Miss | Conf |
| Baseline | 49.25 | 4.46 | 35.41 | 9.38 | 85.38 | 46.67 | **8.03** | 30.68 | 69.22 | 38.36 | 11.45 | **19.41** |
| + E2E VAD | 36.41 | 2.98 | 21.04 | 12.39 | 50.63 | **8.88** | 14.82 | 26.93 | 70.68 | 32.02 | 11.22 | 27.44 |
| + Enhanc | 36.17 | 2.98 | 21.04 | 12.15 | 49.67 | 8.95 | 14.82 | 25.9 | 66.13 | 32.02 | 11.22 | 22.89 |
| + VB reseg | 34.86 | **2.97** | 21.06 | **10.83** | 48.49 | 8.9 | 14.88 | 24.71 | 63.03 | 32.02 | **11.21** | 19.8 |
| + Overlap Assign | **30.76** | 3.73 | **13.64** | 13.39 | **47.49** | 8.9 | 14.88 | **23.71** | **58.59** | **23.08** | 13.15 | 22.36 |

**Table 1**. DER % for different subdatasets and modules.

|  | AMI | | | BabyTrain | | | SRI data | | |
|---|---|---|---|---|---|---|---|---|---|
|  | EER | minDCF | ActDCF | EER | minDCF | ActDCF | EER | minDCF | ActDCF |
| . Baseline with Diar | 17.08 | 0.58 | 0.65 | 14.34 | 0.68 | 0.69 | 21.07 | 0.81 | 0.88 |
| + Enhanc | 16.13 | 0.56 | 0.62 | 10.55 | 0.48 | 0.53 | 19.9 | 0.80 | 0.83 |
| + xvec aug | **12.85** | **0.44** | **0.58** | 9.93 | 0.42 | 0.52 | 16.86 | 0.62 | **0.62** |
| + PLDA aug | 12.89 | 0.44 | 0.58 | **9.28** | **0.37** | **0.6** | **16.37** | 0.63 | 0.63 |

**Table 2**. Detection EER %, minDCF and actDCF for different subdatasets and modules.

as clean data. For the auxiliary network, we choose ResNet-34 x-vector network to filter VoxCeleb set. For the enhancement network, we use a ConvGenNet and a network we design based on CAN.

4. **Voice activity detection(Diar/SPkD)**: Trained on 2s audio chunks for each corpora, with trainable SincNet features (using the configuration [14]). Features are feed into BiLSTM network with binary output (speech/non-speech).

5. **Overlap detection(Diar)**: Follows the same architecture as the VAD where output classes are single/multiple speakers.

6. **Embedding extraction(Diar/SpkD)**: We used an extended TDNN architecture (E-TDNN) presented in [26].

7. **Clustering(Diar)**: the system employs an Agglomerative Hierarchical Clustering (AHC) to compute the speaker labels in a recording. PLDA models were trained on VoxCeleb [25] and adapted to each domain using a small amount of in-domain training data.

8. **Resegmentation and overlap assignment (Diar)**: We first perform resegmentation using HMM-VB resegmentation module. We used the labels from the clustering and the 400 dimensional i-vectors. We use a single VB inference iteration in accordance with [17]. The most likely speaker is assigned to frames detected as speech by the voice activity detector. A second most likely speaker is only assigned for frames detected as overlapped speech.

9. **Traditional speaker detection pipeline**: Includes the speech/feature enhancement, the automatic segmentation obtained from the diarization stage, the shared embedding extractor and PLDA (both with augmentation), and a calibration stage.

We employed Kaldi [27] as our primary toolkit to develop our pipeline. For specific parts, such as VAD and overlap detection we used pyannote [28]. For speech and feature enhancement we implemented the algorithms in a combination of PyTorch and Kaldi.

## 4. RESULTS

We built a robust system that can be useful for the research community[3]. We followed the same pipeline for the corpora, meaning that we achieved some degree of generalization in the models for the VAD and overlap detection, speech and feature enhancement, and the embedding training. The PLDAs are treated separately and are corpus dependant. Diarization provides the first evidence for an effective speaker detection. Table 1 shows the DER by adding a module starting from the single baseline. We can observe improvements for the three sub-datasets. We emphasize the DER relative improvement of 37%, 44% and 15% for AMI, BabyTrain and CHiME5 respectively. One of our main findings was that an enhancement phase, either speech or feature, is necessary when dealing with adverse scenarios. The VB-HMM resegmentation gives some improvement, but the improvement is increased when combined with the overlap detector.

---
[3]Some of the research threads are on-going but with good perspectives of further improvements.

Table 2 presents the results for speaker detection. For the purpose of this paper, we selected a combination of 30s enrollments and above 5s trials. Other combinations are also possible, but they follow the same trend. An automatic segmentation, *i.e.*, an automatic diarization, was performed previously to label the speakers in the recording. We observe the EER relative improvement of 24%, 35% and 22% for AMI, BabyTrain and SRI data. The enhancement provided improvements on the three scenarios. To improve the robustness of the system we used augmentation to train the embedding (x-vector) and for the back-end (PLDA).

## 5. CONCLUSIONS

In this paper we presented our contribution to the JSALT Workshop 2019. We showed which aspects of the pipeline are most influential in challenging scenarios. The very first finding was that speaker detection depends on a previous diarization stage to obtain successful results. Hence, we showed the necessary elements that have to be competitive on their own in the diarization pipeline: VAD, overlap detection, speech/feature enhancement, embedding extractor, clustering with resegmentation, and overlap assignment based on resegmentation. Once we tune this diarization stage, the output can be connected to the detection phase, which is basically a speaker verification system. The speech enhancement, the embedding and PLDA augmentation gave the most improvements. There are open research threads that will require further study such as: customization of speech enhancement for a dataset, exploration of other architectures for feature enhancement, how to handle domain mismatch, unsupervised adaptation in the clustering, and how to highlight the speaker of interest in the detection, among others. As we have shown, the diarization/detection problem is far from being solved in adverse scenarios.